\documentclass{ws-procs9x6}

\begin{document}

\title{Hadron Spectroscopy in electrons-protons Collisions at HERA}

\author{Bob OLIVIER}

\address{DESY, \\
Notkestrasse 85, \\ 
22607 Hamburg, Germany\\ 
E-mail: bob.olivier@desy.de}

\maketitle

\abstracts{
Recent results from H1 and ZEUS on searches for exotic
baryons in $ep$ collisions at HERA are reviewed.
Evidence for the production of the strange pentaquark $\Theta^+$ and of a narrow 
anti-charmed baryon decaying to $D^{*-} p$ together with negative results of pentaquark
searches at HERA are presented.
}

\section{Introduction}

Experimental evidence for an exotic baryon first came in 2003~ from the
observation of a $S=+1$ narrow resonance~\cite{Nakano:2003qx} at $1540 \pm 10$~MeV
which can be associated with an exotic pentaquark
state with content $uudd\bar{s}$. 
Confirmation came quickly from a series of experiments,
with the observation of sharp peaks~\cite{Barmin:2003vv,Stepanyan:2003qr,Barth:2003ja,Asratyan:2003cb} in the $nK^+$ and $pK^0_S$
invariant mass spectrum near $1540$~MeV, in each case with a width limited by the experimental
resolution.

Strong evidence in support of a baryon decuplet comes from the reported
observation of an exotic $S =-2$, $Q =-2$ baryon resonance in $pp$ collisions~\cite{Alt:2003vb}.
A narrow peak at a mass of about $1862$~MeV in the 
$\Xi^-\pi^-$ invariant mass spectrum is proposed as a candidate for the
predicted exotic $\Xi^{--}_{\frac{3}{2}}$ baryon with a  quark content $dsds\bar{u}$.
At the same mass, a peak is observed that is a candidate for $\Xi^{0}_{\frac{3}{2}}$

At HERA electrons (or positrons) of energy $27.5$~GeV are collided with $920$~GeV protons 
providing a center of mass energy $\sqrt{s}$ of $318$~GeV. 
In the following we present the results of the searches
for strange and charmed pentaquarks performed by the 
H1 and ZEUS experiments with an integrated luminosity of up to $126$~pb$^{-1}$
accumulated from 1995 to 2000 during the HERA-I data taking period.

\section{Search for $\Theta^{+}$ and $\Theta^{++}$ }
ZEUS performed a $\Theta^+$ search~\cite{Chekanov:2004kn,spq-zeus-ichep04}
at high energies using the 
$ep$ data taken in the years 1996-2000 with an integrated luminosity of
$121$~pb$^{-1}$. The kinematic region is restricted to the
exchanged boson virtuality domain $Q^2 > 1$~GeV$^2$ and the inelasticity domain $0.01 < y < 0.95$.

The decay chain $\Theta^+ \rightarrow p K^0_S \rightarrow p\pi^+ \pi^-$
has been used. About 866800 $K^0_S$ candidates are selected. 
They are combined with proton candidates selected via the energy-loss measurement
$dE/dx$.

\begin{figure}[htb]
\centerline{
\epsfxsize=4.95cm   
\epsfbox{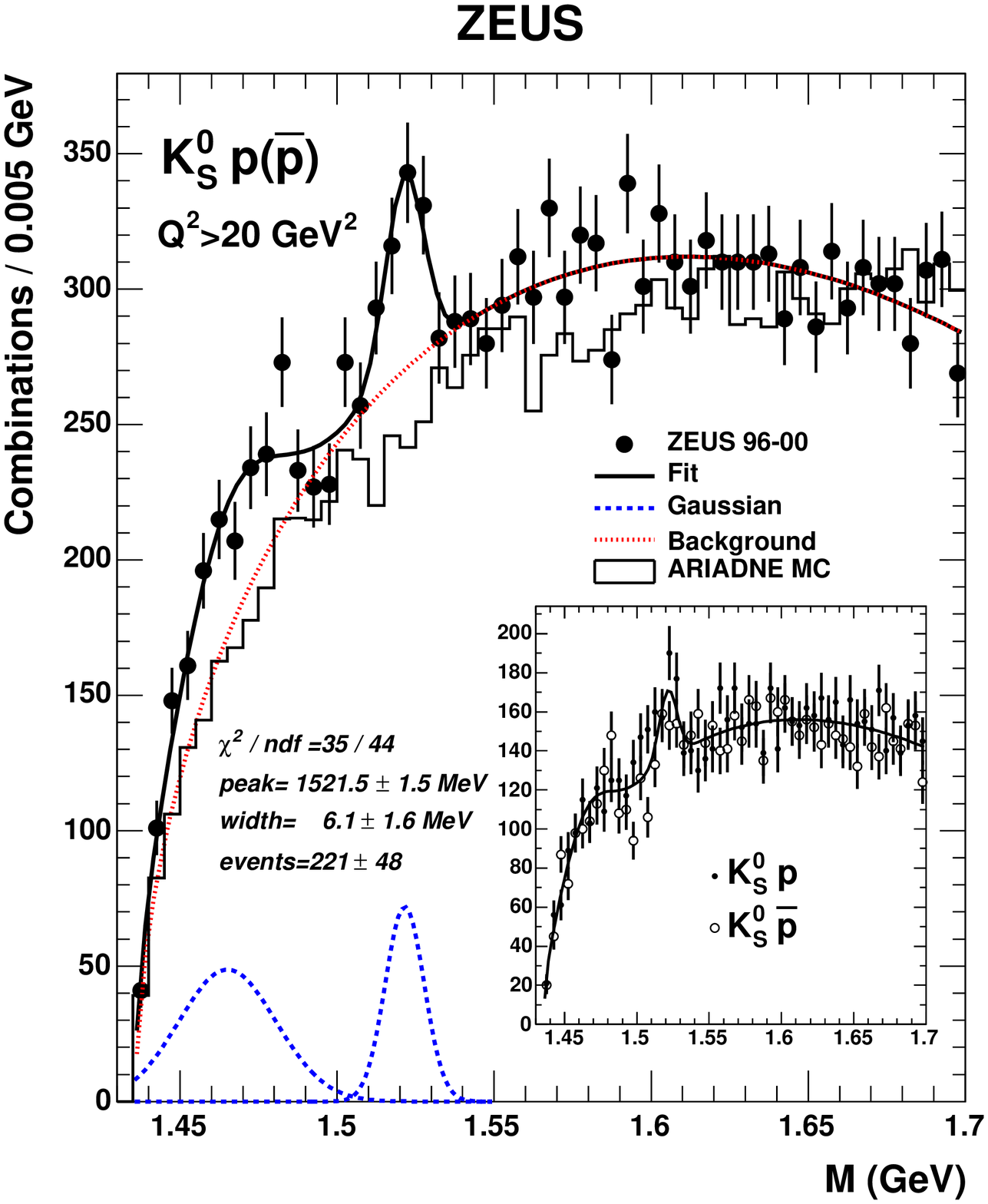}
\hskip 0.3in
\epsfxsize=6.6cm  
\epsfbox{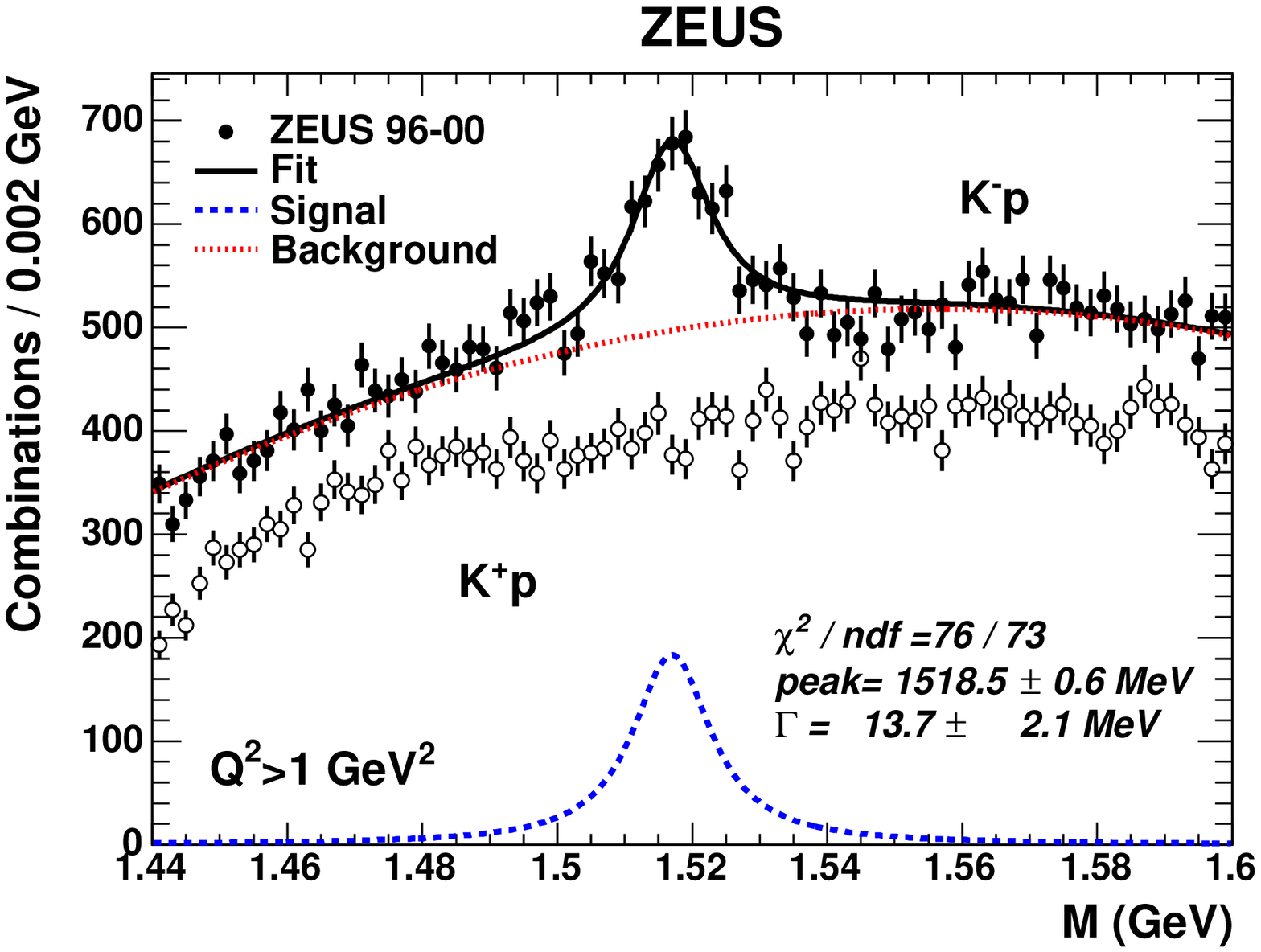}
}
\caption{(left) Invariant  mass distribution $M_ {p\pi^+\pi^-}$ observed by ZEUS
at $Q^2 > 20$~GeV$^2$. 
(right) Invariant mass distribution $M_{pK^+}$ (open dots) and  $M_{pK^-}$
(full dots) observed by ZEUS.}
\label{sPQ}
\end{figure}

The $M_{\pi^+ \pi^-}$ mass distribution shows sign of structure below about $1600$~MeV.
For $Q^2 > 10$~GeV$^2$, a peak is seen in the mass distribution around $1520$~MeV. In
Fig.~\ref{sPQ} the  $M_{\pi^+ \pi^-}$is shown for $Q^2 > 20$~GeV$^2$. The figure includes the Monte
Carlo expectation from ARIADNE~\cite{ariadne} scaled to the data for $M_{\pi^+ \pi^-} > 1650$~MeV.
After scaling ARIADNE does not describe the data at low masses, maybe due to
the absence of the $\Sigma$ bumps in the simulation.
A fit to the data of a smooth background function and two Gaussians, also shown
in Fig.~\ref{sPQ}, gives a signal of $221 \pm 48$ events at a mass of $1521.5 \pm 1.5(stat.)$~MeV
with a significance of $4.6\sigma$. The Gaussian width of $6.1$~MeV is found to be consistent
with the experimental resolution. The signal is observed at similar rate for protons
and for antiprotons suggesting the existence of the anti-pentaquark $\Theta^-$.

ZEUS has also measured the cross section for the production
of the $\Theta^+$  baryons and their antiparticles in the kinematic region $Q^2 > 20$~GeV$^2$,
$0.04 < y < 0.95$, $p_T > 0.5$~GeV and $|\eta| < 1.5$,
$\sigma(ep  \rightarrow e\Theta X \rightarrow e K^0_S p ({\bar p}) X)
= 125 \pm 27(stat.)^{+36}_{-28} (syst.)$ pb.
ZEUS also measured the $\Theta^+$ production cross section 
for higher $Q^2$ thresholds $30$, $40$, and $50$ GeV$^2$.
The $\Theta^+$ cross section  shows no significant dependence on $Q^2$.

After the presentation of this talk  were released results on the $\Theta^+$ search
performed by H1~\cite{spq-h1-dis05} in a similar phase space region than ZEUS. 
No peak was observed in the $M_{\pi^+ \pi^-}$ mass distribution.
At a mass of $1522$~MeV, H1 sets an upper limit on the $\Theta^+$ production 
cross section of the order of $100$~pb. This upper limit is compatible with
the ZEUS observed cross section due to the large error on the latter one.

ZEUS has also searched for the $\Theta^{++}$ signal via its possible decay
$\Theta^{++} \rightarrow K^+ \pi^+$. 
Fig.~\ref{sPQ} shows the $M_{pK^-}$ and $M_{pK^+}$ mass spectra. 
No peak structure is observed in the $M_{pK^+}$
spectrum but in the $M_{pK^-}$ spectrum the well established resonance 
$\Lambda(1520) \rightarrow pK^-$
is clearly seen. As no signal is found in the $\Theta^+$
mass range, this suggests that the $\Theta^+$ could be isoscalar.

\section{Search for $\Xi ^{--}_{\frac{3}{2} }$ and $\Xi^{0}_{\frac{3}{2}}$} 
\begin{figure}[htb]
\centerline{
\epsfxsize=6.6cm   
\epsfbox{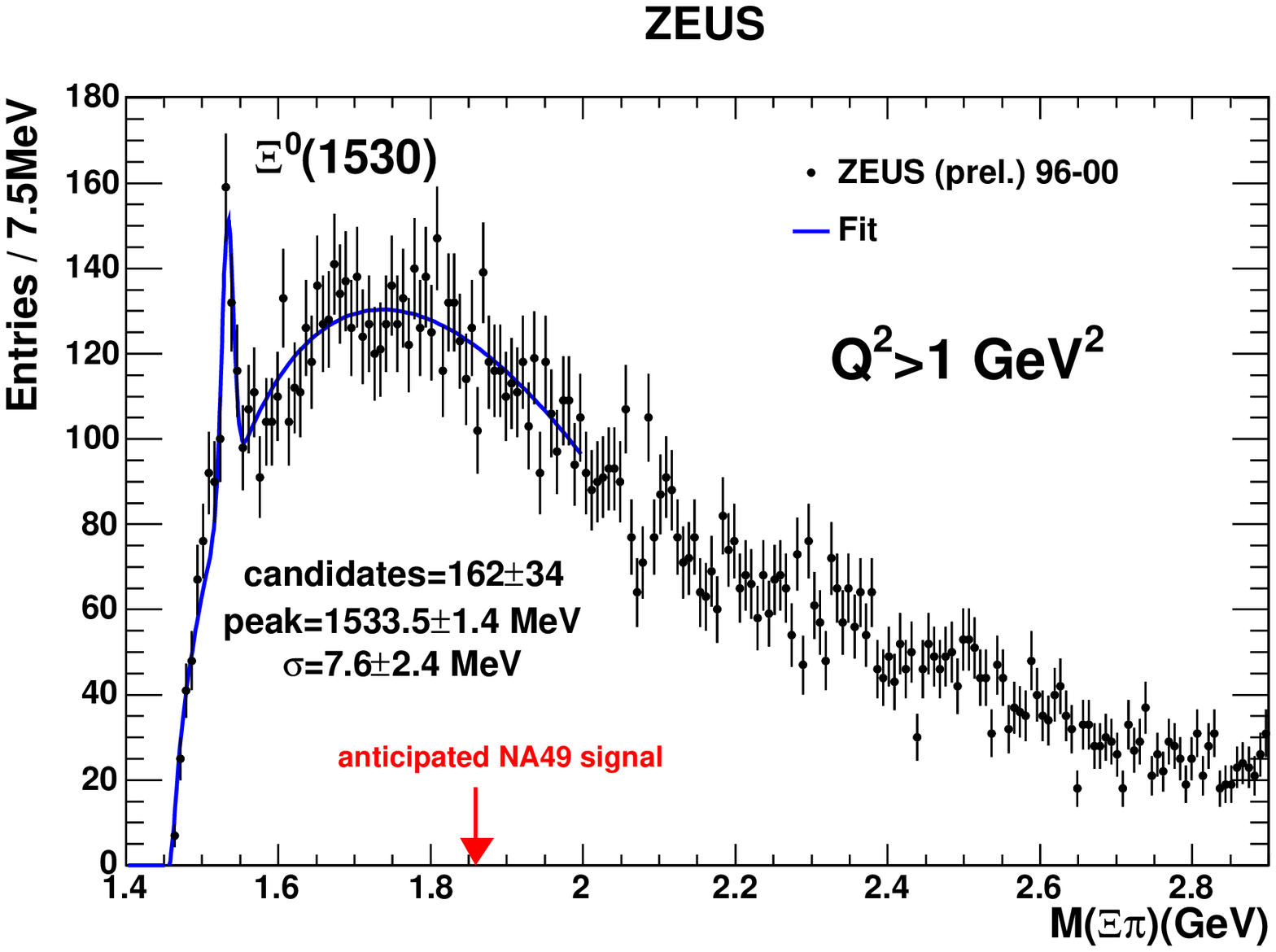}
\hskip 0.3in
\epsfxsize=6.6cm  
\epsfbox{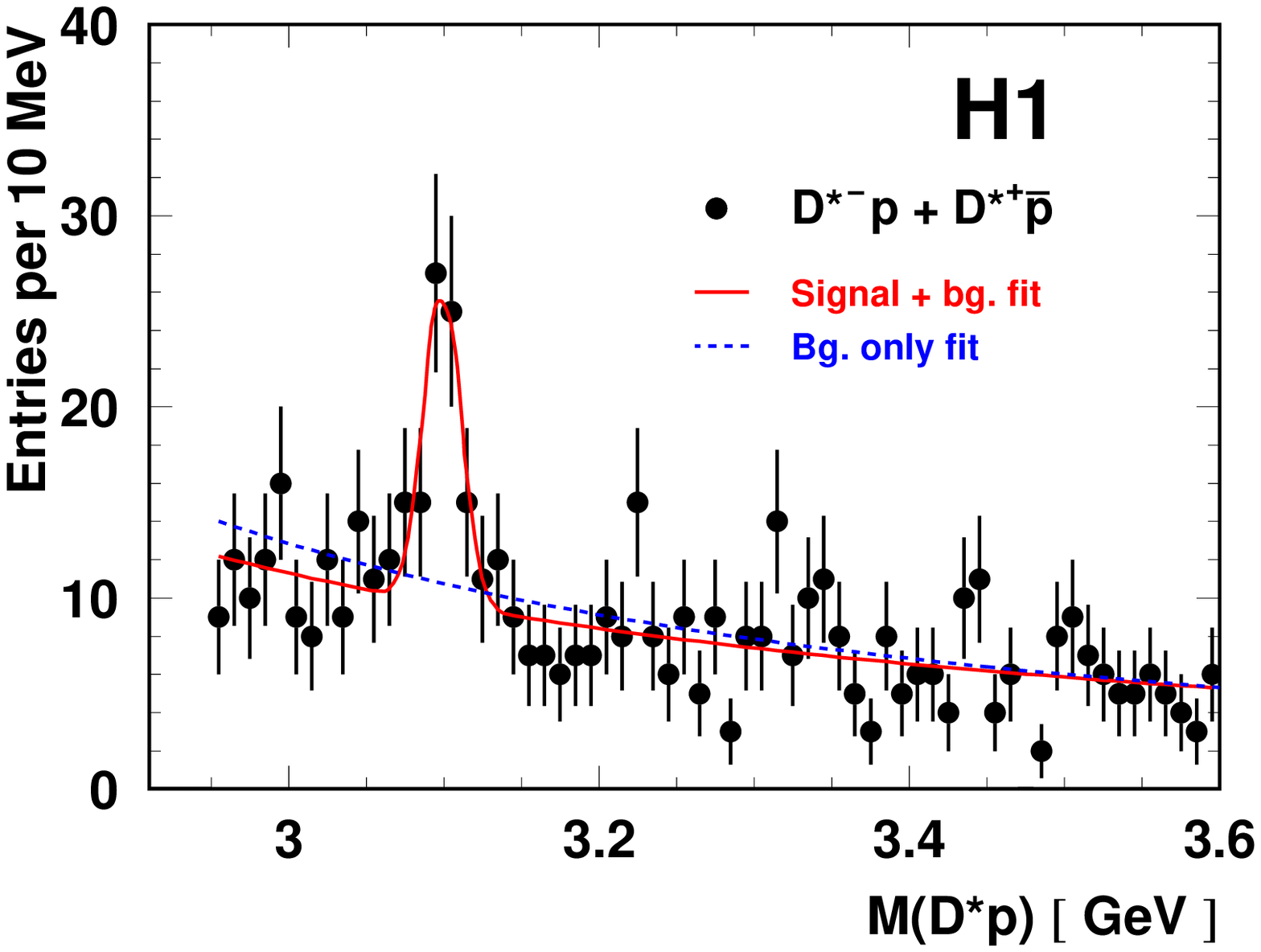}
}
\caption{(left) 
Invariant mass distribution $M_{\Xi\pi}$ observed by ZEUS for $Q^2 > 1$~GeV$^2$ 
and for all four charge combinations combined.
(right) 
Invariant mass distribution $M_{D^* p}$ from H1 for $Q^2 > 1$~GeV$^2$
compared to the fit results with two hypotheses: signal plus background
(solid line) and background only (dashed line).}
\label{cPQ}
\end{figure}

ZEUS has performed an analysis in the channel $\Xi^- \pi^{\pm}$ to search for the strange
pentaquark $\Xi^{--}$ and its neutral partner~\cite{xi-zeus-ichep04}. The decay chain 
$\Xi^{--} \rightarrow \Xi^- \pi^- \rightarrow \Lambda \pi^- \pi^-$ 
has been considered. 
$\Lambda$ baryons were identified by the charged-decay mode, 
$\Lambda \rightarrow p \pi^-$, using pairs of tracks from secondary vertices. These are then combined with
another pion from the primary vertex. Fig.~\ref{cPQ} shows the $M_{\Xi\pi}$ mass distribution
for all possible $\Xi\pi$ charge combinations for $Q^2 > 1$~GeV$^2$ . 
While the $\Xi^0(1530)$ is
clearly visible, no signal is observed around $1860$~MeV as observed by the NA49
collaboration~\cite{Alt:2003vb}. Even when restricting to $Q^2 > 20$~GeV$^2$, 
where the $\Theta^+$ signal was best seen by ZEUS, no signal is observed.
\section{Search for $\Theta_c$}
The production of a charmed pentaquark  $\Theta_c$ has been studied via its decay into
$D^*p$ by H1~\cite{Aktas:2004qf,cpq-h1-dis05} and ZEUS~\cite{Chekanov:2004qm}.

The analysis of H1 is based on the DIS data taken in the years 1996-2000
with a luminosity of $75$~pb$^{-1}$ in the kinematic region $1 <  Q^2 < 100$~GeV$^2$ and
$0.05 < y < 0.7$. The $D^{*\pm}$  charmed meson has been reconstructed via its decay
chain $D^{*+} \rightarrow D^0 \pi^+_S \rightarrow (K^-\pi^+)\pi^+_S$.
Around 3400 $D^*$  candidates are selected,
and are combined with proton candidates selected via $dE/dx$.

The resulting $M_{D^{*-} p}$ distribution in Fig.~\ref{cPQ} shows a clear narrow peak close
to the threshold. 
The signal is both observed in the $D^{*-} p$ and in the $D^{*+}\bar{p}$ sample with compatible mass,
width and rate. 
Log-likelihood fits to the $M_{D^*p}$ distribution are performed. 
The background is parametrised by a power law while a Gaussian is used for the signal.
A signal of $51$ events is observed with a mass of $3099  \pm 3(stat.) \pm 5(syst.)$~MeV
and a width of $12 \pm 3(stat.)$~MeV consistent with the experimental resolution. The
background fluctuation probability has been estimated to be less than
$4\times 10^{-8}$.

A similar search has been performed by ZEUS in both photoproduction and DIS
regimes. Data from the years 1995-2000 with an integrated luminosity of $126$~pb$^{-1}$
have been analyzed. About 9700 $D^*$ candidates are selected for $Q^2 > 1$~GeV$^2$ and
43000 candidates for all data.
No signal is observed at $3.1$~GeV.

Upper limits on the fraction of $D^*$  mesons originating from the $\Theta_c$
decays,
$R = N(\Theta_c \rightarrow D^*p)/N(D^*p)$, were set by ZEUS in the signal window of 
$3.07 <
M_{D^*p} < 3.13$~GeV. This window covers the H1 measurement. The $95\%$ 
confidence level upper limit on the fraction $R$ is $0.23\%$. The upper limit for DIS with
$Q^2 > 1$~GeV$^2$ is $0.35\%$. Thus, the ZEUS results are not compatible
with the report of the H1 collaboration of a charmed pentaquark which contributes
to $1.59\pm 0.33(stat.)^{+0.33}_{-0.45}(syst.)\%$ of the $D^{*\pm}$  
production rate~\cite{cpq-h1-dis05}.

\section{Conclusions}
Recent results from H1 and ZEUS on searches for exotic
baryons in $ep$ collisions at HERA have been presented. 
ZEUS has found evidence for the production of the strange
pentaquark $\Theta^+$. H1 on the contrary has not found any signal compatible
with the $\Theta^+$ and has obtained limits for its production. ZEUS
has not found any evidence for the signal seen by the NA49 collaboration 
attributed to the $\Xi^{--}_{\frac{3}{2}}$. 

H1 has found evidence for the existence of a narrow anti-charmed baryon 
decaying to $D^*p$. This result has not been confirmed by the ZEUS search which
has been performed in a similar kinematic region.

Pentaquark searches are still an open issue.
Of the colliding beam experiments that are currently taking data, the 
HERA experiments H1 and ZEUS are the only ones which have reported the
observation of pentaquark signals to date.
Pentaquark production may be suppressed
in $e^+e^-$ annihilation due to the lack of any particles carrying baryon
number in the initial state colliding beams.
The complicated and high multiplicity hadronic final states
produced in $pp$ and $p \bar{p}$ scattering may obscure any pentaquark
signal, especially if it is dominantly produced at low transverse momentum.
The search for pentaquarks at HERA using the high statistics data from the 
HERA-II data taking period, to be completed in 2007, may thus 
represent a unique opportunity to make progress in the field of exotic 
hadron spectroscopy.


\begin{thebibliography}{0}
\bibitem{Nakano:2003qx}
T.~Nakano {\it et al.}  [LEPS Collaboration],
Phys.\ Rev.\ Lett.\  {\bf 91}, 012002 (2003).

\bibitem{Barmin:2003vv}
  V.~V.~Barmin {\it et al.}  [DIANA Collab.],
  Phys.\ Atom.\ Nucl.\  {\bf 66}, 1715 (2003).

\bibitem{Stepanyan:2003qr}
  S.~Stepanyan {\it et al.}  [CLAS Collaboration],
  Phys.\ Rev.\ Lett.\  {\bf 91}, 252001 (2003);
  V.~Kubarovsky {\it et al.},  {\it ibid.}
  {\bf 92}, 032001 (2004).

\bibitem{Barth:2003ja}
  J.~Barth {\it et al.}  [SAPHIR Collaboration],
  Phys.\ Lett.\ B {\bf 572}, 127 (2003).

\bibitem{Asratyan:2003cb}
  A.~E.~Asratyan {\it et al.},
  Phys.\ Atom.\ Nucl.\  {\bf 67}, 682 (2004).

\bibitem{Alt:2003vb}
  C.~Alt {\it et al.}  [NA49 Collaboration],
  Phys.\ Rev.\ Lett.\  {\bf 92}, 042003 (2004).

\bibitem{Chekanov:2004kn}
  S.~Chekanov {\it et al.}  [ZEUS Collaboration],
  Phys.\ Lett.\ B {\bf 591}, 7 (2004).

\bibitem{spq-zeus-ichep04} ZEUS Collaboration, Contributed paper to ICHEP04, Beijing, China, Abstract 10-0273 (2004).

\bibitem{spq-h1-dis05} H1 Collaboration, Contributed paper to DIS05, Madison, USA.

\bibitem{xi-zeus-ichep04} 
ZEUS Collaboration, Contributed paper to ICHEP04, Beijing, China, Abstract 10-0293 (2004).

\bibitem{ariadne} L.~Lonnblad, Comput. Phys. Commun. {\bf 71}, 15 (1992).

\bibitem{Aktas:2004qf}
  A.~Aktas {\it et al.}  [H1 Collaboration],
  Phys.\ Lett.\ B {\bf 588}, 17 (2004).


\bibitem{cpq-h1-dis05} H1 Collaboration, Contributed paper to DIS05, Madison, USA.

\bibitem{Chekanov:2004qm}
  S.~Chekanov {\it et al.}  [ZEUS Collaboration],
  Eur.\ Phys.\ J.\ C {\bf 38}, 29 (2004).
\end{thebibliography}
\end{document}